\documentclass[aps,prl,twocolumn,superscriptaddress,showpacs]{revtex4}

\usepackage{graphicx}

\begin{document}

\title{Electron Beam Instability in Left-Handed Media}

\author{Yury P. Bliokh}
\affiliation{Frontier Research System, The Institute of Physical and Chemical Research (RIKEN),
Wako-shi, Saitama 351-0198, Japan} \affiliation{Department of Physics, Technion-Israel Institute of
Technology, Haifa 32000, Israel}
\author{Sergey Savel'ev}
\affiliation{Frontier Research System, The Institute of Physical and Chemical Research (RIKEN),
Wako-shi, Saitama 351-0198, Japan} \affiliation{ Department of Physics, Loughborough University,
Loughborough LE11 3TU, United Kingdom}
\author{Franco Nori}
\affiliation{Frontier Research System, The Institute of Physical and Chemical Research (RIKEN),
Wako-shi, Saitama 351-0198, Japan} \affiliation{Center for Theoretical Physics, Department of
Physics, CSCS, University of Michigan, Ann Arbor, Michigan 48109-1040, USA}

\begin{abstract}
We predict that two electron beams can develop an instability when passing through a slab of left-handed media (LHM). This instability, which is inherent only for LHM, originates from the backward Cherenkov radiation and results in a self-modulation of the beams and radiation of electromagnetic waves. These waves leave the sample via the rear surface  of the slab (the beam injection plane) and form two shifted bright circles centered at the beams.  A simulated spectrum of radiation has well-separated lines on top of a broad continuous spectrum, which indicates dynamical chaos in the system.
The radiation intensity and its spectrum can be controlled either by the beams' current or by the distance between the two beams.   
\end{abstract}


\maketitle

Metamaterials (also known as left-handed media (LHM) \cite{Veselago}),
characterized by having negative both permittivity $\varepsilon$ and permeability $\mu$, show numerous remarkable counterintuitive features 
\cite{LHM}, including: negative refraction \cite{Veselago}, inverted Doppler shift and reversed Cherenkov cone (backward Cherenkov radiation) \cite{Pafomov}, as well as surface waves propagating  along the interface between right- and left-handed media \cite{Ruppin}. These properties (promising for applications in a wide frequency range for subwavelength imaging and lensing \cite{Pendry2}, subwavelength \cite{Engheda} and open \cite{Notomi} resonators \cite{Bliokh2}, nonradiating  configurations \cite{Boardman}, etc.) have stimulated enormous scientific activity during the past decade. However, as far as we know, the \textit{collective} interaction of charged particle beams with electromagnetic waves in LHM has not been studied. Nevertheless, one can expect a very nontrivial interaction among particles in the beams since the Cherenkov radiation emitted by a particle propagates backward \cite{Pafomov,LHMCherenkov}, producing strong positive distributed feedback for particles moving behind. Such a strong coupling can create an instability and chaotic (i.e., very irregular) motion in the beams and waves. In this article we predict the instability of electron beams in LHM associated with tunable self-sustained electromagnetic radiation.

\begin{figure}[tbh]
\centering \scalebox{0.35}{\includegraphics{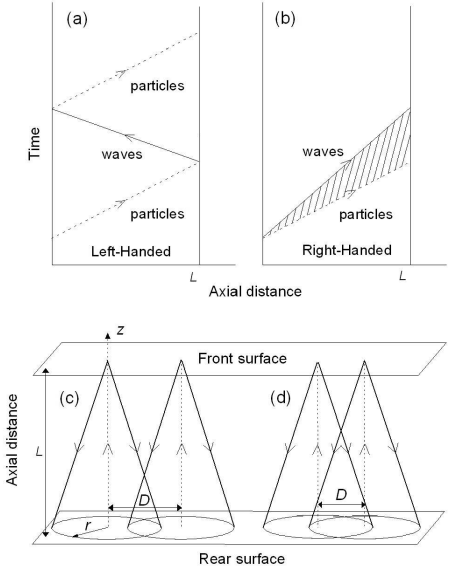}} 
\caption{Waves (solid line) and particles (dashed lines) characteristics in LHM (a) and RHM (b) slabs with thickness $L$. For LHM, particles and waves transport information in the opposite direction, while, for RHM, information can  only be transmitted forward. (c,d) Cherenkov field, emitted by particles (dashed lines), located within Cherenkov cones shown by solid lines. For  case (c) the distance between the beams is too large, $D>D_{{\rm max}}$, and the radiation emitted by one beam cannot reach the other beam, so there is no interaction between beams. For  case (d), $D<D_{{\rm max}}$, the radiation emitted by one beam reaches the tail of the other beam, creating a particle-density modulation producing a strong beam coupling. \label{Fig1}}
\end{figure}

In order to provide an intuitive picture of the effect discussed below, we compare information transport by electromagnetic waves (solid line) and particles (dashed line), shown in Fig.~\ref{Fig1}, for slabs of left-handed (a) and right handed (b) media. Information about any perturbation created, for instance, at the rear (left) surface is transfered either along the wave characteristics (solid line) and/or along the particle trajectories (dashed line) deep towards the sample. For right-handed media (RHM), both information fluxes are directed forward and perturbations propagate within the shaded region between the characteristics (Fig.~\ref{Fig1}b). Thus, any knowledge about the perturbation leaves the sample in the forward direction after a finite time. In contrast, for LHM, particles and waves transport information in opposite directions. Therefore, the information transported from the rear to the front surfaces by particles is returned back by the emitted waves (Fig.~\ref{Fig1}a). Simultaneously, these ``returning'' waves perturb the particles entering the sample later on (in the rear surface), and the  process of information transport  will oscillate back and forth, and never stops. Similar processes occur in microwave devices as traveling wave tubes (TWT) (similar to a slab of RHM) and backward wave oscillators (BWO) (similar to a slab of LHM) \cite{Gilmour}.
 
It is well known \cite{Gilmour} that positive feedback in BWO produces an electron beam instability and self-excited microwave radiation. Thus, a similar instability can be expected in LHM. Consider, for example, a small perturbation (denoted here as ``bunch'') of the beam density with a longitudinal dimension  smaller than the wave-length of the synchronous wave (i.e., a wave with phase velocity  equal to the beam particles' velocity). Particles forming the bunch radiate \textit{coherently} and the radiated electromagnetic field is not compensated (due to destructive interference) by other beam particles. This wave acts on electrons moving behind the bunch and periodically modulates their velocities. These velocity variations  modulate  the beam density resulting in secondary bunches  in the beam. Due to the periodicity of the emitted waves, radiation produced by the bunches sum up \textit{coherently} and the total field increases. A stronger field leads to a faster and deeper modulation of the beam density that, in  turn, strengthens the field. As a result, the amplitude of the synchronous wave grows exponentially. This is the so-called beam instability, which has been discovered simultaneously in plasma physics \cite{Akhiezer,Bohm} and microwave electronics \cite{Pierce}. Thus, two collective effects (the coherent Cherenkov radiation of the particles forming the bunch and the coherence between the bunches) constitute the physical basis of the beam instability.

Here we consider two separated beams moving along a strong magnetic field and interacting via their Cherenkov radiations propagating backward in 3D LHM and predict a beam instability. This instability develops when the distance between the beams is within a certain interval and when the beam currents exceed a certain threshold.

\textit{Model}. --- We describe LHM by a frequency-dependent permittivity and permeability (see, e.g., \cite{Pendry2,Ziolkovski,Ranmakrishna,Boardman}): 
 \begin{equation}\label{eq1}
\varepsilon(\omega)=\mu(\omega)=1-{\omega^2_p\over\omega^2+i\omega\nu},
\end{equation}
where $\nu$ is the collision frequency and $\omega_p$ is the plasma frequency. The medium is ``left-handed'' when the wave frequency $\omega<\omega_p$.

\begin{figure}[tbh]
\centering \scalebox{0.47}{\includegraphics{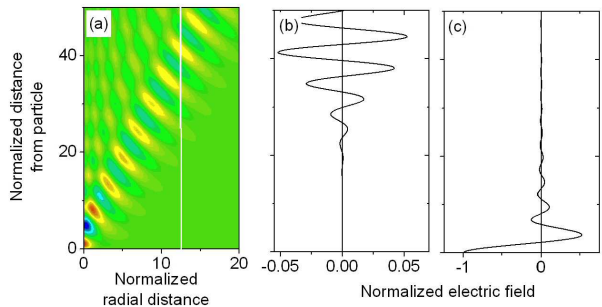}} \caption{(Color online) Spatial distribution of the  Cherenkov radiation emitted by a single particle. (a) The field distribution $E_z$ in the $(\rho,\;\xi^\prime)$-plane, where $\rho=k_{0z}r$ and $\xi^\prime=k_{0z}(z-vt)$ are the normalized radial  and axial distances from the particle; $k_{0z}$ is the axial component of the wave vector $\vec{k_0}$ corresponding to the maximum of the spectrum of radiation. The electric field amplitude is depicted with the factor  $\sqrt{\rho}$, which compensates the field decay $E\sim\rho^{-1/2}$. (b), (c): axial distribution of the field $E_z$ for $\rho=12.5$ (white line in Fig.~2a) and $\rho=0$ accordingly. Note that $E_z(\xi)$ for $\rho=12.5$ has the form of a wave packet with a distinguishable periodicity (b), whereas $E_z(\xi)$ for $\rho=0$ is fast-decaying and non-periodic (c). \label{Fig2}}
\end{figure}

We consider two parallel beams, separated by a distance $D$ propagating along the $z$-direction (the limit $D\rightarrow 0$, corresponding to one beam, is discussed below).  Charged particles in these beams emit Cherenkov radiation with the $z$-component of the electric field described by the standard formula \cite{Jelly} 
\begin{eqnarray}\label{eq2}
E_{z0}={e\over c^2}\intop\left({1\over n^2\beta^2}-1\right)J_0\left({\omega\over
v}\sqrt{n^2\beta^2-1}r\right)\times\nonumber\\
\times\cos\left[{\omega\over v}(z-vt)\right]\omega\;d\omega,
\end{eqnarray}  
where $v$ is the particle velocity, $\beta\equiv v/c$, and $n=-\sqrt{\varepsilon\mu}$ is the refraction index. The integration domain in a LHM is restricted by the conditions $\omega\geq 0$, $|n|\beta>1$, $\varepsilon<0$ and $\mu<0$ [i.e., $0\leq\omega\leq\omega_p\beta/(1-\beta)$]. The other components of the radiated fields will not be needed because  only the $E_z$-component governs the particle motion  along the $z$-axis in a strong guiding magnetic field. Below we consider the non-relativistic limit, $\beta\ll1$, and small dissipation, $\nu\ll\beta^{1/2}\omega_p$.
The motion of particles interacting through their emitted waves can be described by:
\begin{equation}\label{eq3}
{d^2z_i^{(1,2)}\over dt^2}={e\over m}\sum_j\left[E_{ij}^{(1,2)}+E_{ij}^{(2,1)}\right],
\end{equation}
where $z_i^{(k)}$ is the coordinate of the $i$-th particle in the $k$-th beam, $k=1,\;2$, and   $E_{ij}^{(k)}=E_{z0}[(z_i^{(\ell)}-z_{jr}^{(k)})-\beta c(t-t_{jr}^{(k)}),r_{ij}]$ is the radiated field of the $j$-th particle from the $k$-th beam acting on the $i$-th particle from the $\ell$-th beam, $z_{jr}^{(k)}$ and $t_{jr}^{(k)}$ are the position and the time when the $j$-th particle radiated the wave that reaches the $i$-th particle at position $z_i$ at time $t$, $r_{ij}=0$ when $k=\ell$ and $r_{ij}=d$ otherwise. $E_{z0}(z-\beta c t,\;r)$ is the ``elementary'' field with space-time structure defined by Eq.~(\ref{eq2}) and depicted in Fig.~\ref{Fig2}a. 

The essential property of  the Cherenkov radiation emitted by a single particle radiation is that the field is mostly concentrated along the line $\rho=\xi\tan\alpha$ and the angle $\alpha=\arcsin(1/3)$ is independent of the particle velocity; in full analogy with the wake that a moving ship produces on the surface of the sea  \cite{Lighthill}. This means that we can associate a unique group velocity $v_g$ to the largest fraction of the radiation. Thus, we can approximate the position $z_{jr}$ and the time $t_{jr}$ as:
\begin{equation}\label{eq4}
t_{jr}=t_{j0}+\intop_0^{z_{jr}}{dz^\prime\over v_j(z^\prime)},\hspace{3mm} 
z_{jr}=z+|v_{gz}|\left(t-t_{jr}\right)\cos\alpha,
\end{equation}
where $t_{j0}$ is the time when the particle crosses the rear surface $z=0$ of the LHM slab. 

Hereafter the following dimensionless variables are used: $\xi=k_{0z}z$, $\eta=k_{0z}x$, $\zeta=k_{0z}y$, and $\tau=k_{0z}v_0t$ ($v_0$ is the unperturbed beam velocity). 

\begin{figure}[tbh]
\centering \scalebox{0.42}{\includegraphics{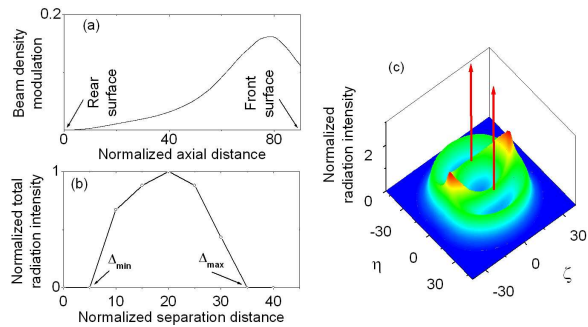}}
\caption{(Color online) (a)  Amplitude of the beam density modulation (due to bunching of particles) along the beam, divided by the average density. The maximum of the density modulation occurs near the front surface; thus, beams radiate with higher intensity near this surface. (b) Total radiation intensity $I_{\rm tot}$ versus the separation distance $\Delta$ between two beams. This intensity is nonzero for $\Delta_{\rm min}<\Delta<\Delta_{\rm max}$. (c) Spatial distribution of the radiation intensity $I(\eta,\zeta)$ at the rear surface of the LHM slab. This distribution has the form of two shifted circles centered at the beams.  \label{Fig3}}
\end{figure}

\textit{Instability of two beams.} --- We simulated the motion of electrons in two thin beams, separated by a distance $D$ in the transverse direction, using Eqs.~(\ref{eq3}) and (\ref{eq4}); each simulation takes  time $\tau_s\gg \mathcal{L}$, where $\mathcal{L}=k_{0z}L$ is the dimensionless thickness  of the slab. 
The simulations showed that, at certain conditions, a beam instability develops: any small perturbation develops into a strong modulation of the beam density. Inside the slab, the beam modulation  increases strongly away from the rear surface and exhibits a maximum approaching the front surface (Fig.~\ref{Fig3}a). The beam density modulation is the source of radiation which is transported along the reverse Cherenkov cone (see Fig.~\ref{Fig2}a  and Figs.~\ref{Fig1}c,d) by the backward waves to the rear surface.  The total field $E_z$ is calculated as a sum of the ``elementary'' fields radiated by all the particles inside the slab. The total radiation intensity, $I_{{\rm tot}}=\intop\; d\eta d\zeta\;  I(\eta,\zeta,\xi=0)= \tau_s^{-1}\intop d\eta d\zeta d\tau \; E^2_z(\xi=0,\eta,\zeta,\tau)$ is not zero when the beam separation $\Delta=k_{0z}D$ is within a certain interval $\Delta\in (\Delta_{{\rm min}}, \Delta_{{\rm max}})$ (Fig.~\ref{Fig3}b).  The window $(\Delta_{{\rm min}}, \Delta_{{\rm max}})$ of this instability weakly depends on the beam current (for simplicity, we consider two beams with the same current $J_b$). The spatial distribution of the radiation intensity $I(\eta,\zeta)$ at the rear surface forms two overlapping or intersecting rings, as  shown in Fig.~\ref{Fig3}c. These bright rings could be seen by an observer located outside the slab. 

The physical origin of the suppression of the instability for any large separation distance $\Delta>\Delta_{{\rm max}}$ is rather obvious: indeed, in this case the neighboring beams are located outside their corresponding Cherenkov cones and cannot interact via their emitted waves (Fig.~\ref{Fig1}c). At small distances $\Delta<\Delta_{{\rm min}}$, the instability is suppressed due to three different reasons. First, the electromagnetic field emitted by a particle decreases fast (see Fig.~\ref{Fig2}c) along the beam. Second, the spatial period of the decaying wave varies behind the particle, which destroys the spatial periodicity of the beam modulation and, as a result, suppresses the coherence of the radiation emitted by this modulation. These two reasons are common for the destruction of the beam instability in both 3D LHM and BWO  and explain why there is no instability in the one-beam system.
The other, third reason, is unique for 3D LHM: when the distance between the beams decreases, the region of intersection of one beam with the Cherenkov cone formed by the second beam is shifted toward the front surface of the slab. The beam modulation is amplified along the distance between this intersection and the front surface of the slab. This distance is shortened when the beam separation decreases and becomes insufficiently long for an instability to develop. Particles in the two different beams interact effectively and the radiation intensity reaches its maximum when the intersection between the beam and the Cherenkov cone occurs near the front surface (Fig.~\ref{Fig1}d). This is in agreement with our simulations (Fig.~\ref{Fig3}a). Therefore the distance between the beams can effectively control the intensity of the radiation.  

\begin{figure}[tbh]
\centering \scalebox{0.95}{\includegraphics{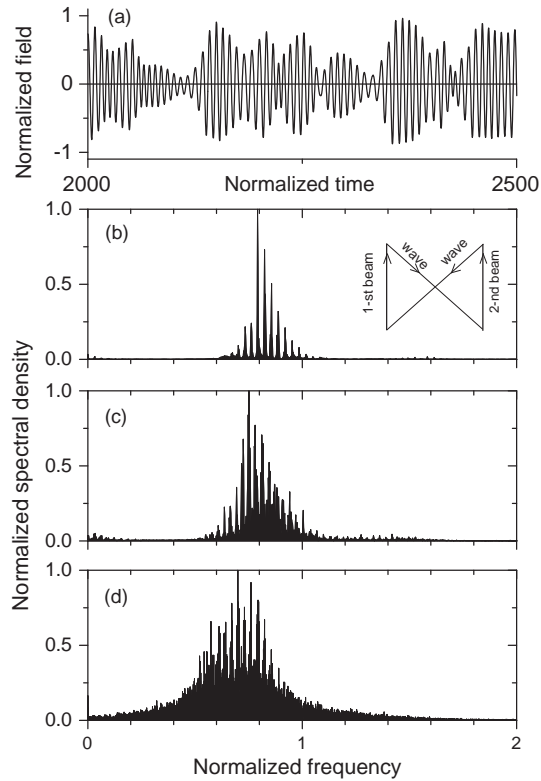}}
\caption{(a)  Time-dependence of the electric field $E(\tau)$ of the emitted wave at the rear surface of the LHM slab. The irregular modulation of the wave amplitude indicates chaotic dynamics of the beams. (b-d) Radiation spectra of a two-beam system when the beam current grows. Spectra show well separated narrow lines on top of a continuous background. The current increases by a factor $1.5$ from (b) to (d). \label{Fig4}}
\end{figure}

\textit{Properties of radiation.} --- Let us consider the radiation  for the optimal separation between the beams. An analogy with BWO \cite{Gilmour} points out that we can expect three different radiation regimes: (i) no radiation for either a weak  beam current or small slab thickness, (ii) monochromatic radiation above a certain current/thickness thresholds, and (iii) periodic or chaotic self-modulation at larger currents/thickness. Indeed, we found no instability at either small current or thin samples. Above a threshold we always observed a chaotic self-modulation. An example of the time-dependence of the electric field of the emitted wave at the rear surface is presented in Fig.~\ref{Fig4}a. This dependence clearly shows the irregular modulation of the wave amplitude. The spectra of the wave is shown in Fig.~\ref{Fig4} (b-d), for increasing beam current. All of these three presented spectra  have a similar structure: narrow spectral lines whose overlap forms a continuous spectrum. With increasing current, the lines become broader and the strength of the continuous spectrum grows, which is usual for microwave devices with a so-called ``frequency'' scenario of transition to chaos \cite{Bliokh}. 
The frequency gap between the spectral lines is determined by the time needed for a signal to close the feedback loop (signal is transported first by one beam, then the signal is carried back by the electromagnetic wave from the beam head to the tail of the second beam, then forward along the second beam, and afterward by the wave from the head of the second beam to the initial point; see inset in Fig.~\ref{Fig4}b). Therefore, the distance between the spectral lines can be tuned by the distance between the beams or can be designed by a proper choice of the slab thickness.

\textit{Conclusions.} --- We predict a beam instability in 3D left-handed media. This instability produces strong radiation which can be tuned either by the beam current or by geometrical parameters. The spectrum of the radiation has well separated lines on top of a broad background. Since now LHM can be fabricated \cite{Grigorenko} in the optical frequency range, the predicted effect can be used to generate and amplify optical irregular signals. Further extension of the analogy between traditional microwave devices and left-handed media will allow new insights for new potential applications of LHM.      
\begin{acknowledgments}
We acknowledge partial support from the NSA, LPS, ARO, NSF grant No. EIA-0130383, JSPS-RFBR
06-02-91200, MEXT Grant-in-Aid No. 18740224, the EPSRC via No. EP/D072581/1, EP/F005482/1,
the AQDJJ ESF network-program, and the JSPS CTC Program.
\end{acknowledgments}

\end{document}